\documentclass[twocolumn,showpacs,prl,amsmath,amssymb]{revtex4-2}
\usepackage{xcolor,epsfig,bm,amsmath,amssymb}
\newcommand{\cN}{{\cal N}} 
\newcommand{\cO}{{\cal O}}

\newcommand{\cS}{{\cal S}}

\newcommand{\E}      {{\mathbb{E}}}                 
\newcommand{\R}      {{\mathbb{R}}}                 
\newcommand{\Z}      {{\mathbb{Z}}}                 

\newcommand{\vx} {{\bm x}}

\newcommand{\nn}{\nonumber} 
\newcommand{\be}{\begin{equation}} 
\newcommand{\ee}{\end{equation}}  
\newcommand{\bea}{\begin{eqnarray}}
\newcommand{\eea}{\end{eqnarray}}
\newcommand{\barr}{\begin{array}}
\newcommand{\earr}{\end{array}}
\newcommand{\bcent}{\begin{center}} 
\newcommand{\ecent}{\end{center}}  
\newcommand{\bit}{\begin{itemize}}
\newcommand{\eit}{\end{itemize}}
\newcommand{\bq}{\begin{quote}}
\newcommand{\eq}{\end{quote}}

\begin{document}
\title{Level-Set Percolation of Gaussian Random Fields on Complex Networks}
\author{Reimer K\"uhn}
\affiliation{Mathematics Department, King's College London, Strand, London WC2R 2LS,UK}
\date{\today}


\begin{abstract}
We provide an explicit solution of the problem of level-set percolation for multivariate Gaussians defined in terms of weighted graph Laplacians on complex networks. The solution requires an analysis of the heterogeneous micro-structure of the percolation problem, i.e., a self-consistent determination of locally varying percolation probabilities.  This is achieved using a cavity or message passing approach. It can be evaluated, both for single large instances of locally tree-like graphs, and in the thermodynamic limit of random graphs of finite mean degree in the configuration model class. 
\end{abstract}

\maketitle
{\em Introduction.}---At its core, percolation describes a geometric phase transition, at which as a function of the relative density of existing links in a lattice or a graph --- either by construction or after random removal of a subset of edges or vertices --- the system either decomposes into a collection of finite clusters of contiguously connected vertices, or on the contrary exhibits a so-called giant connected component (GCC) that occupies a finite fraction of vertices in the large system limit\cite{Kes82}. Apart from an intrinsic interest in percolation transitions and the critical singularities associated with them, they are of relevance in many other contexts. E.g., below the percolation threshold, where a system conists of a collection of finite isolated clusters, diffusive or hopping transport via edges of a graph is clearly impossible, and such systems would therefore be insulators. In a different context, the size of an SIR epidemic can be mapped on the size of its GCC in a perolcation problem where edge retention probabilities are given by probabilities of transmitting a disease before recovery \cite{Grassb83, New02, More+02, KarrerNewm10}. This connection has been exploited to formulate effective vaccination strategies for diseases spreading through contact networks (e.g., \cite{PastorVesp02, CHbA03,CPH+08}). More generally, percolation has been studied to assess the robustness of complex networks, both natural and artificial, against random failures of their components or against targeted attacks (e.g., \cite{CCN++00,AJB00,ACB10}). At a more fundamental level, a non-percolating system cannot support stable phases with spontaneous macroscopic long range order at any non-zero temperature (see for example \cite{Sti83}), and it has been argued that gene regulatory networks would for the same reasons not be able to support multi-cellular life, if they were composed of small independent clusters of interacting genes \cite{Hannam+19,TKA20}. 

There is a version of the percolation problem which is of a radically different nature than the case of independent Bernoulli percolation just described. It is concerned with the distribution of sizes of contiguous clusters, over which a random field exceeds a given level $h$. It has been studied over continuous spaces ($\R^d$) \cite{MoSt83a, MoSt86, MuiVan20,CaoSan21, MuiSev22, Mui22, MuRiVa23}, over lattices ($\Z^d$) \cite{MoSt83b, BLM87, RoSz13, DrPrRo18, Mui22}, or over random graphs \cite{AbSz18, AbCe20a, AbCe20b, DrPrRo22, CoKe23, DrPrRo23, Cer23, CeLo23}. Because of correlations between values of a random field at different points in space --- in important cases in fact {\em long range\/} correlations --- the problem is much harder than that of independent Bernoulli percolation. Indeed, early on Molchanov and Stepanov \cite{MoSt83a,MoSt83b} disproved the na\"ive intuition according to which there  would {\em always\/} exist a finite critical level $h_c$, above which all contiguous clusters for which the random field exceeds the level $h>h_c$ would be bounded, whereas for $h < h_c$ there would be an extensive giant connected component (referred to as $h$GCC in what follows) over which the random field exceeds the level $h$. While a range of important results about level-set percolation have been obtained over the years, including existence (e.g. \cite{MoSt83a,MoSt83b,MoSt86,BLM87, RoSz13,AbSz18,Mui22,MuRiVa23}) and sharpness \cite{Mui22} of the percolation transition, uniqueness (e.g. \cite{AbCe20b}) and extensivity of the $h$GCC (e.g. \cite{CoKe23,Cer23,CeLo23}) in the percolating phase, as well as critical exponents of the transition on transient graphs (e.g. \cite{CeLo23,DrPrRo23}), significant gaps remain. E.g., despite recent intense activity and considerable progress in the study of level-set percolation on random graphs \cite{AbSz18, AbCe20a, AbCe20b, DrPrRo22, CoKe23, DrPrRo23, Cer23, CeLo23}, explicit solutions of this problem that go beyond a characterization of the near critical regime on regular trees do to the best our our knowledge still elude us, in marked contrast to the case of Bernoulli percolation on random graphs, for which full solutions are meanwhile textbook material (e.g., \cite{AB02, DM03, NewBk10}). It is the purpose of the present letter to fill this gap.

{\em Level-Set Percolation for Gaussian Free Fields on Random Graphs.}---We consider a (random) graph of $N$ vertices on which a multivariate Gaussian field is defined via
\be 
P(\vx) =\frac{1}{Z}\,\exp\Big(-\cS(\vx)\Big)\ ,
\label{GFF}
\ee 
in which $\cS(\vx)$ quadratic in the $x_i$, i.e.,
\be
\cS(\vx) = \frac{1}{2}\mu\sum_i x_i^2+\frac{1}{4}\sum_{i,j} K_{ij}(x_i-x_j)^2
\label{H}
\ee
with $\mu \ge 0$ and $K_{ij}=K_{ji} >0$ for vertices of the network connected by an edge. For $\mu  = 0$ the field is referred to as massless. For the analysis of level-set percolation on random graphs, it turns out to be essential to be able to characterize its heterogeneous micro-structure, i.e. the node dependent probabilites of vertices in the graph to belong to the $h$GCC. This can be done by adapting an approach developed in \cite{KNZ14,KuRog17}. It is based on cavity or message passing ideas specifically designed to analyse problems on locally tree-like graphhs. As, we are only interested in heterogeneous percolation probabilities, a somewhat simpler version outlined, e.g., in \cite{ShirKaba10, BTB+21} can be used. 

For a node $i$ to belong to the $h$GCC, the Gaussian field at $i$ must itself exceed the specified level, i.e.,  $x_i\ge h$, {\em and\/} it must be connected to the $h$GCC via at least one of its neighbours. Introducing indicator variables $n_i\in \{0,1\}$ which signify whether $i$ is ($n_i=1$) or is not ($n_i=0$) in the  $h$GCC, we require
\be 
n_i = \chi_{\{x_i\ge h\}}\,\Big(1 - \prod_{j\in \partial i} (1 - \chi_{\{x_j\ge h\}} n_j^{(i)}\big)\Big)\ ,
\label{ni}
\ee 
in which the first factor expresses the fact that the Gaussian field at $i$ must itself exceed the specified level, i.e.,  $x_i\ge h$, while the second factor expresses the fact that $i$ is connected to the hGGC via at least one neighbour. This requires that for at least one $j\in \partial_i$ the Gaussian field must exceed the specified level ($x_j \ge h$), {\em and\/} it must be connected to the $h$GCC via one of {\em its\/} neighbours other than $i$ (on the cavity graph $G^{(i)}$ from which $i$ and the edges connected to it are removed); this is expressed by the cavity indicator variable $n_j^{(i)}$  taking the value $n_j^{(i)}=1$.

For the cavity indicator variable $n_j^{(i)}$ to take the value 1, it is required that on $G^{(i)}$ the node $j$ is itself connected to the $h$GCC via at least one of its neighbours other than $i$. This entails that
\be
n_j^{(i)} = 1 - \prod_{\ell\in \partial j\setminus i} \big(1- \chi_{\{x_\ell\ge h\}} n_\ell^{(j)}\big)\ .
\label{nji}
\ee
Averaging Eqs.\,\eqref{ni} and \eqref{nji} over possible realizations of the Gaussian field $\vx$ with joint PDF described by Eqs.\,\eqref{GFF} and \eqref{H} is facilitated by the fact that --- conditioned on $x_i$ --- the averages over the $\big(\chi_{\{x_j\ge h\}} n_j^{(i)}\big)_{j\in \partial i}$ in Eqs.\,\eqref{ni} factor in $j$, if the graph in question is a tree, and that such factorization becomes asymptotically exact on locally tree-like graphs in the thermodynamic limit. Analogous factorization is possible for averages over the $\big(\chi_{\{x_\ell\ge h\}} n_\ell^{(j)}\big)_{\ell \in \partial j\setminus i}$ in Eqs.\,\eqref{nji}, when conditioned on $x_j$.

Performing the average over the Gaussian field $\vx$ in this way, we obtain $g_i = \E_\vx[n_i] = \E_{x_i}\Big[\E_\vx[n_i|x_i]\Big]$ from Eq.\,\eqref{ni} as
\be
g_i\!\!=\!\!\E_{x_i}\Bigg[\chi_{\{x_i\ge h\}}\bigg(\!1\! - \!\prod_{j\in \partial i}\!\!
        \Big(1 - \E_\vx\big[\chi_{\{x_j\ge h\}} n_j^{(i)}\big| x_i\big]\Big)\bigg)\Bigg]
\ee
by factorization of conditional expectations. Then, using further conditioning, we have 
\bea
\E_\vx\Big[\chi_{\{x_j\ge h\}} n_j^{(i)}\big| x_i\Big]\!\!
                    &=&\!\! \E_{x_j}\Big[\chi_{\{x_j\ge h\}} \big| x_i\Big]\nn\\
                    & &\!\! \times \E_\vx\Big[n_j^{(i)}\big| \{x_j\ge h\}, x_i\Big]\nn\\
                    &=&\!\! H_j(h|x_i)~g_j^{(i)}
\eea
where we have introduced 
\be
H_j(h|x_i) = \E_{x_j}\Big[\chi_{\{x_j\ge h\}}\Big| x_i\Big]
\label{Hji}
\ee
and 
\be 
g_j^{(i)} = \E_\vx\Big[n_j^{(i)}\big|\{x_j\ge h\}\Big]\ .
\label{gjidef}
\ee
In Eq.\,\eqref{gjidef} we have used the fact that for conditional expectations of observables such as $n_j^{(i)}$ pertaining to the cavity graph $\E_\vx\big[n_j^{(i)}\big| \{x_j\ge h\}, x_i\big]=\E_\vx\big[n_j^{(i)}\big| \{x_j\ge h\}\big]$. Putting things together, we have
\be
g_i\!\! =\!\! \rho_i^h 
 \Bigg(\! 1\! -\!\E_{x_i}\bigg[\!\prod_{j\in \partial i}\!\! \Big(\! 1\! -\! H_j(h|x_i) g_j^{(i)}\Big)\Big|\{x_i\ge h\}\bigg]
\Bigg)
\label{gi}
\ee
with $\rho_i^h = \E_{x_i}[\chi_{\{x_i\ge h\}}]$. 

Following an entirely analogous line of reasoning and using the same sequence of conditionings, we can evaluate the $g_j^{(i)}$ defined in Eq.\,\eqref{gjidef} by evaluating the conditional average of $n_j^{(i)}$ using Eq.\,\eqref{nji}, giving
\be
g_j^{(i)}\! =\! 1 - \E_{x_j}\bigg[\!\prod_{\ell\in \partial j\setminus i}\!\!\Big(\! 1\!-\! H_\ell(h|x_j) g_\ell^{(j)}\Big)\Big|\{x_j\ge h\}\bigg]\ .
\label{gji}
\ee
Equations \eqref{gi} and \eqref{gji} for the $g_i$ and the $g_j^{(i)}$ can be evaluated, once single-site marginals and joint densities on adjacent sites of the Gaussian field defined by Eqs.\,\eqref{GFF} and \eqref{H} are known; the latter are required to evaluate the conditional  probabilites $H_j(h|x_i)$ defined in \eqref{Hji} (and similarly the $H_\ell(h|x_j)$ appearing in Eq.\,\eqref{gji}), while the former are required to evaluate $x_i$ expectations in Eqs.\,\eqref{gi} and $x_j$ expectations in Eqs.\,\eqref{gji}, respectively. They are obtained by their own cavity type analysis, which has in fact been performed in \cite{Ku+07} for single-site marginals of harmonically coupled systems on random graphs, and in \cite{Ku08,Rog+08} in the context of the spectral problem of sparse symmetric random matrices. All that is needed are the (Gaussian) single-site marginals $P_i(x_i)$ of $P(\vx)$, as well as the corresponding single-site cavity margininals $P_j^{(i)}(x_j)$ for $j\in\partial i$ on the cavity graph $G^{(i)}$, in terms of which joint densities on adjacent sites are easily obtained. Key identities needed in the analysis are reproduced in the supplementary material \cite{KuSupMat24} to this letter.  Single-site marginals and single-node cavity marginals are fully characterized by their inverse variances (or precisions) $\omega_i$  and $\omega_j^{(i)}$, respectively. The latter are obtained by solving the the system (4) of cavity self-consistency equations in \cite{KuSupMat24}. The $H_j(h|x_i)$ and the $H_\ell(h|x_j)$ can be expressed in closed form in terms of error functions, but the conditional $x_i$-expectation of the product in Eq.\,\eqref{gi} and similarly the conditional $x_j$-expectation of the product in Eq.\,\eqref{gji} will have to be evaluated numerically.

With all ingredients fully defined, Eqs.\,\eqref{gji} constitute a set of coupled self-consistency equation for the $g_j^{(i)}$. They can be solved iteratively at given level $h$ on large instances of locally tree-like (random) graphs, starting from random initial conditions. Using the solutions, one obtains the node-dependent percolation probabilities $g_i$ from Eqs.\,\eqref{gi}.

The value of the percolation threshold follows from a linear stability analysis of Eqs.\,\eqref{gji}. They are {\em always\/} solved by $g_j^{(i)} \equiv 0$. This solution becomes unstable, indicating the percolation transition, where the the largest eigenvalue of the Hessian of the r.h.s. of Eq.\,\eqref{gji} evaluated at $g_j^{(i)} \equiv 0$ exceeds 1. The Hessian is a weighted version of a so-called non-backtracking matrix, with non-zero elements
\be
B_{(ij),(j\ell)} =  \E_{x_j}\big[H_\ell(h|x_j)\Big|\{x_j\ge h\}\big]
\label{nb}
\ee
for $j\in \partial i$ and $\ell\in  \partial j\setminus i$, and $B_{(ij),(k\ell)} = 0$ otherwise. Performing an appropriately adapted weakly non-linear expansion of Eqs.\,\eqref{gji} as in \cite{KuRog17}, one obtains site dependent percolation probabilities to linear order in $h_c-h$ as
\be
g_i \simeq \alpha ~ (h_c-h) \sum_{j\in \partial i} v_j^{(i)}
\ee
where $\bm v = \big(v_j^{(i)}\big)$ is the Frobenius right eigenvector of corresponding to the largest eigenvalue $\lambda_{\rm max}(B)\big|_{h=h_c}= 1$ of the non-backtracking matrix \eqref{nb} evaluated at $h_c$, normalized s.t. $||v||_1=1$, and $\alpha$ is an amplitude given in Eq.\,(23) of the supplementary material, which also includes a derivation of the $\cO((h_c-h)^2)$ contribution to the $g_i$ \cite{KuSupMat24}.

\begin{figure}[h]
  \includegraphics[width = 0.45\textwidth]{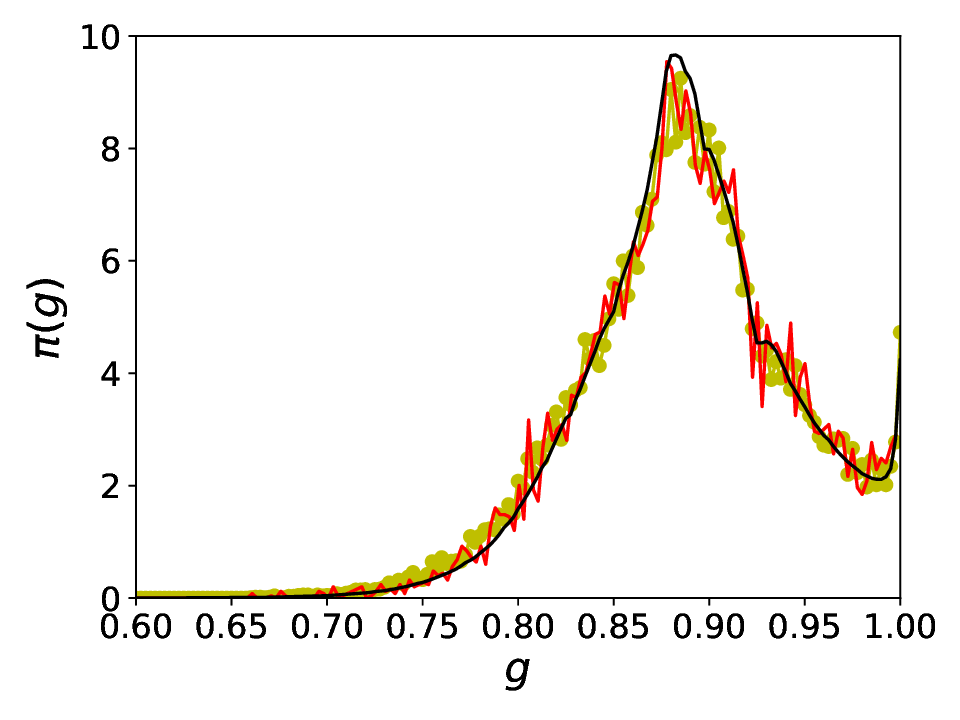}\\
  \caption{(color online) Distribution $\pi(g)$ or local percolation probabilities for a random graph with power law degree distribution $k \sim k^{-3}$ for $2 \le k \le 100$ at $h=-1$ and $\mu=0.1$. We compare {\bf (i)} results of a numerical simulation of a single instance of a graph of $N=50,000$ vertices, averaging over 5,000 realizations of Gaussian field configurations to obtain the PDF of the $g_i$ (yellow dots), with {\bf (ii)} results of a single instance cavity analysis for a graph of the same size (red full line), and {\bf (iii)} the result of an analysis in the thermodynamic limit (black full line).}
  \label{FigER3pw3m2}
\end{figure}

{\em Thermodynamic Limit.}---For random graphs in the configuration model class, i.e., the class of gaphs that are maximally random subject to a given degree distribution $p_k= \mbox{Prob}(k_i=k)$, one can analyse the level-set percolation problem in the thermodynamic limit of infinite system size. Assuming that a limiting probability law for the joint distribution of the cavity probabilities $g_j^{(i)}$ and the cavity precisions $\omega_j^{(i)}$ exists, probabilistic self-consistency compatible with the self-consistency equations \eqref{gji} for the $g_j^{(i)}$ and with Eqs.\,(4) of the supplementary material \cite{KuSupMat24} for the $\omega_j^{(i)}$ entails a self-consistency equation for the joint PDF $\tilde\pi(\tilde g,\tilde \omega)$, which is documented as Eq.\,(12) of the supplementary material. That equation is efficiently solved by a population dynamics algorithm. The limiting PDF of local percolation probabilities is then evaluated from its solution.

In Fig.\,\ref{FigER3pw3m2}, we present an example of a distribution of level-set percolation probabilities for a system with a fat-tailed degree distribution, which shows that the theoretical analyses agree very well with a numerical simulation. Simulations are, of course, affected by sampling flucutations and by finite size effects (creating details depending on the specific single realization of the generated random graph), while the single instance cavity analysis is only affected by finite size effects. Further results, both for different systems and a range of values for the level $h$, can be found in the supplementary material \cite{KuSupMat24}. Remarkably, as also documented in \cite{KuSupMat24}, for a massless Gaussian field the marginal node dependent precisions $\omega_i$ turn out to be a very precise, although not exact, predictor for the node dependent percolation probabilities $g_i = g_i(h)$ at a given level $h$, which appears to be independent of the graph type. This is particularly interesting as the $g_i$ are {\em much\/} harder to evaluate than the $\omega_i$.  However that almost perfect correlation is lost for fields with non-zero mass $\mu>0$. 

{\em Random Regular Graphs.}---Specializing to random regular graphs (RRGs) with uniform couplings, more explicit results can be obtained. The key observation is that in the thermodynamic limit all nodes and all edges of the system are equivalent. Hence the self-consistency equation for the uniform cavity precisions on a RRG of degree $c$ (or $c$RRG) reads
\be
\tilde\omega = \mu + (c-1) \frac{K\tilde\omega}{K+ \tilde\omega}\ .
\label{tomRRG}
\ee
This equation is solved by
\be
\tilde\omega_\pm\! =\!\frac{1}{2}\Big[\mu+ K(\!c-\!2) \pm \sqrt{[\mu\!+\!K(\!c-\!2)]^2\! +\! 4K\mu}\Big]\ ,      
\ee
the relevant (physical) solution being $\tilde\omega=\tilde\omega_+$. This entails a self-consistency equation for the uniform cavity percolation probabilities $g_j^{(i)}\equiv \tilde g$ of the form
\be
\tilde g = 1 - \E_{x}\Big[\Big(1 - H(h|x)\,\tilde g\Big)^{c-1}\Big|\{x\ge h\}\Big]\ ,
\label{tg}
\ee
in which $H(h|x)$ is a conditional expectation of the type defined in Eq.\,\eqref{Hji}, evaluated on the $c$RRG, and $x\sim \cN(0,1/\omega)$, with
\be
\omega = \mu + c \frac{K\tilde\omega}{K+ \tilde\omega} 
\ee
the uniform single-site precision on the $c$RRG. Equation \eqref{tg} is a simple scalar equation for $\tilde g$ which is easily solved numerically. It always has the trivial solution $\tilde g =0$, which becomes unstable below a critical value $h_c$ of the level $h$ which follows from a linear stability analysis of Eq.\,\eqref{tg} and is given as the solution of
\be
(c-1) \E_{x}\Big[H(h|x)\Big|\{x\ge h\} \Big] = 1\ .
\label{hc}
\ee
\begin{figure}[h]
  \includegraphics[width = 0.45\textwidth]{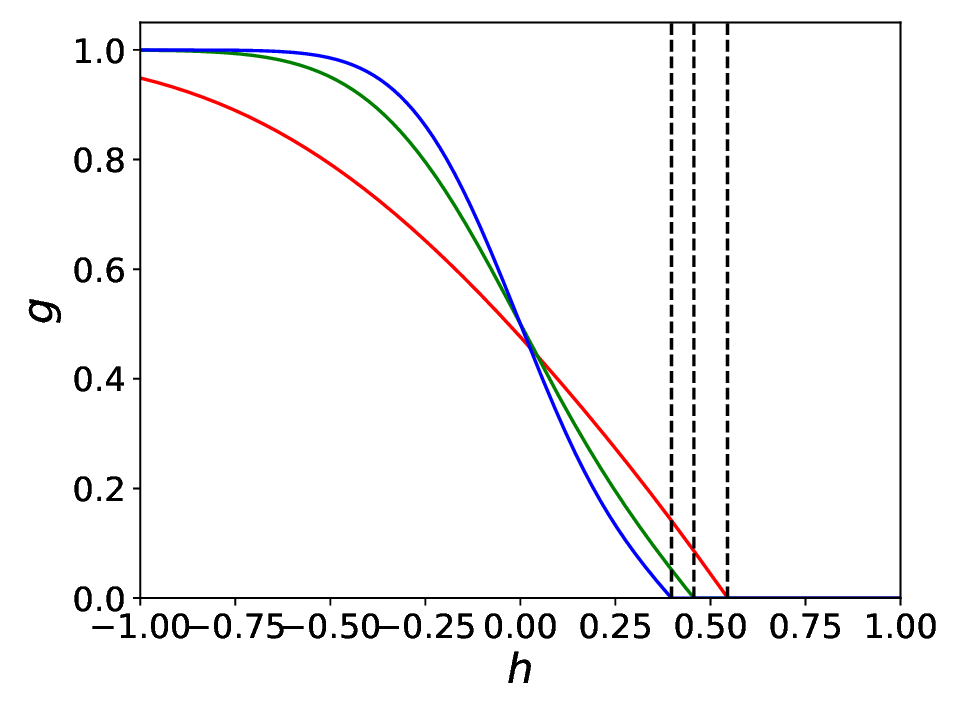}\\
  \caption{(color online) Percolation probability $g$ as a function of the level $h$ for $c$RRGs with $c = 4$, 12, and 20. The steepness of the curves increases with $c$. Critical levels $h_c$ as obtained from Eq.\,\eqref{hc} are indicated as vertical dashed lines. For the three values of $c$ shown here, they decrease with increasing $c$, and they agree perfectly with results of a numerical solution of Eq.\,\eqref{tg}.}
  \label{FigRRG4-12-20}
\end{figure}

From the solution of Eq.\,\eqref{tg} at $h< h_c$ one obtains
\be
g = \rho^h \Big(1 - \E_{x}\Big[\Big(1 - H(h|x)\,\tilde g\Big)^{c}\Big|\{x\ge h\} \Big]\Big)\ ,
\ee
with $\rho^h = \E_{x}[\chi_{\{x\ge h\}}]$, as the value of the percolation probability at level $h$. Percolation probabilities thus computed as functions of the level $h$  are shown in Fig.\,\ref{FigRRG4-12-20} for $c$RRGs with uniform  $K=1$ and $\mu=0$ for three different values of $c\ge 4$. For the range of $c$ values shown critical percolation thresholds $h_c$ are decreasing with increasing $c$, but, as shown in Fig.\, 2 of the supplementary material, there is non-monotonicity of $h_c$ as a function of $c$ in the range $c\in\{3,4,5\}$.

{\em Summary and Discussion.}---In this paper we presented an explicit solution of the problem of level-set percolation of Gaussian free fields on locally tree-like random graphs, both for finite large instances and in the thermodynamic limit of infinite system size for random graphs in the configuration model class with finite mean degree. The former requires the simultaneous solution of a set self-consistency equations for locally varying single-node cavity percolation probabilities $g_j^{(i)}$ and for the locally varying single node cavity precisions $\omega_j^{(i)}$. The latter instead requires solving a nonlinear integral equation for their joint PDF $\tilde\pi(\tilde g,\tilde \omega)$. Though we have restricted ourselves to a uniform mass parameter $\mu$, such a restriction is not a matter of principle and can easily be relaxed. We found our results to be in excellent agreement with simulations. Simplifications are possible in the case of RRGs for which the uniform single-node cavity percolation probabilities $g_j^{(i)}\equiv \tilde g$ are obtained as solutions of a single scalar equation, from which in turn a single scalar equation for the uniform percolation probability $g$ is easily derived. While our methods are non-rigorous, they are expected to be exact in the large system limit, a fact that should be amenable to rigorous proof. It is hoped that some the methods and heuristics used in the present paper could be useful for the analysis of wider classes of level-set percolation problems.  

It would be interesting to investigate, whether the approach of \cite{KuRog17} which is {\em also\/} capable of giving distributions of the sizes of {\em finite\/} clusters, both in the non-percolating and in the percolating phase can be carried over to the present case of level-set percolation. A second as yet unsolved problem concerns the stability analysis of the integral equation (12) of the supplementary material, which could in principle allow one to obtain critical percolation levels $h_c$ for configuration model networks directly in the thermodynamic limit. We hope to address some of these open problems in the near future.

{\em Acknowledgements.}---It is a pleasure to thank Guillaume Conchon-Kerjan for bringing this problem to my attention and for explaining his earlier results.
\bibliography{GLSPerc.bbl}
\end{document}


\title{Supplementary Material --- Level-Set Percolation of Gaussian Random Fields on Complex Networks}
\author{Reimer K\"uhn\\
Mathematics Department, King's College London, Strand, London WC2R 2LS,UK}
\date{\today}
\maketitle
              
\begin{abstract}
These notes contain supplementary material for the letter. We provide the key Gaussian identities needed for the evaluation of the cavity equations and formulate the probabilistic self-consistency equation for the joint probability distribution of cavity percolation probabilities and cavity precisions in the thermodynamic limit for random graphs in the configuration model class. We report examples of level dependent distributions of percolation probabilities for an Erd\H{o}s-R\'enyi (ER) graph and for a graph with a power-law degree distribution, as well as a rather surprising finding relating local level-set percolation probabilities with local marginal precisions for the case of massless Gaussian fields. Moreover, we evaluate degree dependent critical thresholds $h_c$ for random regular graphs. We also provide an asymptotic analysis of the general single instance cavity equations in the vicinity of the percolation threshold $h_c$.
\end{abstract}

\section{Gaussian Identities}
 The averages and conditional averages of indicator functions appearing in the theory require the evaluation of expectations over node-dependent single-site marginals of the multivariate Gaussian defined by Eqs.\,(1), (2) of the letter, as well as averages over conditional distributions of such Gaussians, conditioned w.r.t. the value of the multivariate Gaussian on a neighbouring site.
 
 Their evaluation uses cavity type reasoning of the same form used to compute single-node marginals and self-consistency equations for single-node cavity marginals as used before in the context of the theory of harmonically coupled systems on graphs \cite{Ku+07} or the theory of sparse random matrix spectra \cite{Ku08,Rog+08}. We collect key identities here.

For a free multivariate Gaussian field on a graph with joint Gaussian density given by Eqs.\,(1), (2) of the letter, all marginals are themselves Gaussian, and of the form
\be
P_i(x_i) = \frac{1}{Z_i} \exp\Big(-\frac{1}{2} \omega_i x_i^2\Big)  
\label{GMarg}
\ee
with $Z_i = \sqrt{2\pi/\omega_i}$ and $\omega_i$ denoting their precisions (inverse variances). On a tree, and approximately on a locally tree-like graph, we have
\be
P_i(x_i) \propto \exp\Big(-\frac{1}{2} \mu x_i^2\Big) \prod_{j\in\partial i} \int \rd x_j\,\exp\Big(-\frac{1}{2} K_{ij}(x_i-x_j)^2\Big) P_j^{(i)}(x_j)\ ,
\label{MargCav}
\ee
in which the $P_j^{(i)}(x_j)$ are the marginals of the $x_j$, $j\in\partial i$, on the cavity graph $G^{(i)}$. Given that the cavity marginals must themselves be Gaussian, and denoting by $\omega_j^{(i)}$ their precisions, the $x_j$ integrals in Eq.\,\eqref{MargCav} entail that the marginal precisions $\omega_i$ are given by \cite{Ku+07, Ku08, Rog+08}
\be
\omega_i = \mu + \sum_{j\in\partial i} \frac{K_{ij} \omega_j^{(i)}}{K_{ij}+ \omega_j^{(i)}}\ ,
\label{Prec}
\ee
with the $\omega_j^{(i)}$ still to be determined. Following an analogous line of reasoning for the $P_j^{(i)}(x_j)$ , one can conclude that the $\omega_j^{(i)}$ must satisfy the self-consistency equations
\be
\omega_j^{(i)} = \mu + \sum_{\ell\in\partial j\setminus i} \frac{K_{j\ell} \omega_\ell^{(j)}}{K_{j\ell}+ \omega_\ell^{(j)}}\ .   
\label{CavPrec}
\ee
Equations \eqref{Prec} and \eqref{CavPrec} become asymptotically exact on locally tree like graphs in the thermodynamic limit.

It is straightforward to demonstrate --- again following the same line of reasoning --- that bi-variate Gaussian marginals for two adjacent nodes on the graph are of the form
\be
P_{ij}(x_i,x_j) = \frac{1}{Z_{ij}} \exp\Big(-\frac{1}{2} K_{ij}(x_i-x_j)^2 - \frac{1}{2} \omega_i^{(j)} x_i^2 - \frac{1}{2} \omega_j^{(i)} x_j^2\Big)\ ,
\label{Pij}
\ee
with
\be
Z_{ij} = \sqrt{\frac{(2\pi)^2}{(K_{ij}+\omega_i^{(j)})(K_{ij}+\omega_j^{(i)}) - K_{ij}^2}}
\ee
by normalization. It is then an elementary computation to obtain conditional distributions from the joint pdf Eq.\,\eqref{Pij},
\be
P_{j}(x_j|x_i)  = \frac{1}{\sqrt{\frac{2\pi}{K_{ij} +\omega_j^{(i)}}}}\,
\exp\bigg(-\frac{1}{2} \big(K_{ij}+\omega_j^{(i)}\big) \Big(x_j -\frac{K_{ij} x_i}{K_{ij} +\omega_j^{(i)}}\Big)^2\bigg)\ .  
\ee
The above results allow one to evaluate
\be
\rho_i^h = \E_{x_i}[\chi_{\{x_i\ge h\}}] =  H(\sqrt{\omega_i} h)
\ee
and similarly
\be
H_j(h|x_i) = \E_{x_j}\Big[\chi_{\{x_j\ge h\}}\Big| x_i\Big] =  H\bigg(\sqrt{K_{ij} + \omega_j^{(i)}}\,\Big(h - \frac{K_{ij} x_i}{K_{} + \omega_j^{(i)}} \Big) \bigg)\ ,
\label{Hji}
\ee
where
\be
H(z) = \int_z^\infty \frac{\rd x}{\sqrt{2\pi}} \, \exp\Big( -\frac{1}{2} x^2\Big) = \frac{1}{2} \mbox{erfc}\big(z/\sqrt 2\big)\ .
\ee
With these results, all ingredients of the self-consistency equations (10) of the letter as well as expressions for the local percolation probabilities $g_i$ given by Eq.\,(9) of the letter are well defined once a solution to Eqs.\,\eqref{CavPrec} for the $\omega_j^{(i)}$ has been obtained.

\section{Thermodynamic Limit}
We now proceed to analyse Gaussian level-set percolation in the thermodynamic limit of infinite system size for random graphs in the configuration model class, i.e., for the class of graphs that are maximally random subject to a given degree distribution $p_k= \mbox{Prob}(k_i=k)$. We take their mean degree $\langle k\rangle = \sum_k k p_k$ to be finite.

Assuming that a probability law for the joint distribution of the cavity probabilities $g_j^{(i)}$ and the cavity precisions $\omega_j^{(i)}$ with probability density $\tilde\pi(\tilde g,\tilde \omega)$ in the thermodynamic limit exists, probabilistic self-consistency simultaneously compatible with self-consistency equations \,\eqref{CavPrec} above for the $\omega_j^{(i)}$, {\em and\/} with the self-consistency equations (10) of the letter for the $g_j^{(i)}$, i.e., with
\be
g_j^{(i)}\! =\! 1 - \E_{x_j}\bigg[\!\prod_{\ell\in \partial j\setminus i}\!\!\Big(\! 1\!-\! H_\ell(h|x_j) g_\ell^{(j)}\Big)\Big|\{x_j\ge h\}\bigg]\ .
\label{gji}
\ee
allows one to obtain the PDF $\tilde\pi(\tilde g,\tilde \omega)$ as follows. One averages the right hand sides of Eqs.\,\eqref{CavPrec} and \eqref{gji} over all realizations for which $\omega_j^{(i)} \in (\tilde \omega, \tilde \omega+\rd \tilde \omega]$ and  $g_j^{(i)} \in (\tilde g, \tilde g+\rd \tilde g]$ to obtain (see e.g. \cite{KuRog17} for a similar line of reasoning)
\be
\tilde\pi(\tilde g,\tilde \omega)\! =\! \sum_k \frac{k}{\langle k\rangle}p_k\! \int\! \prod_{\ell=1}^{k} \rd \tilde \pi(\tilde g_\ell,\tilde\omega_\ell)\,
\left\langle\!\delta\Big(\tilde\omega\! -\!\Omega_{k-1}\Big) \delta\bigg(\tilde g\! -\! \Big(1\!- \!
\E_{x}\Big[\prod_{\ell=1}^{k-1}\big(1\!-\! H_\ell(h|x)\,\tilde g_\ell\big)\big|\{x\ge h\}\Big]\Big) \bigg)\right\rangle_{\{K_\ell\}}
\label{tpi}
\ee
in which 
\be
\Omega_{q} = \Omega_{q}(\{\tilde\omega_\ell\}) = \mu + \sum_{\ell=1}^q \frac{K_\ell \tilde\omega_\ell}{K_\ell +\tilde\omega_\ell}\ ,
\ee
and
\be 
H_\ell(h|x) = \E_{x_\ell}\Big[\chi_{\{x_\ell\ge h\}}\Big|x \Big] =  H\bigg(\sqrt{K_\ell + \tilde \omega_\ell}\,\Big(h - \frac{K_{\ell} x}{K_{\ell} +\tilde\omega_\ell} \Big) \bigg)\ ,
\label{Hell}
\ee
is the probability that the Gaussian field on the $\ell$-th node adjacent to a node with Gaussian field $x$ (to which it is coupled via $K_\ell$) does itself exceed the value $h$. In Eq.\eqref{tpi}, $\frac{k}{\langle k\rangle}p_k$  is the probability for a random neighbour of a node to have degree $k$, and the expectation w.r.t. $x$ is evaluated for $x \sim \cN(0,1/\Omega_k)$, while $\langle \dots \rangle_{\{K_\ell\}}$ denotes an average over the $K_\ell$. Moreover, we have introduced the shorthand $\rd \tilde \pi(\tilde g_\ell,\tilde\omega_\ell) = \rd \tilde g_\ell \rd \tilde\omega_\ell\, \tilde \pi(\tilde g_\ell,\tilde\omega_\ell)$. Equation \eqref{tpi} is very efficiently solved by a population dynamics algorithm. From the solution we obtain
\bea 
\pi(g,\omega)\!\! &=&\!\! \sum_k\! p_k\! \int\!\! \prod_{\ell=1}^k \rd \tilde \pi(\tilde g_\ell,\tilde\omega_\ell)\left\langle\!\delta\Big(\omega\! -\!\Omega_{k}\Big) \delta\Big(g\! -\! \E_{x}[\chi_{x \ge h}] \Big(1- \E_{x}\Big[\prod_{\ell=1}^{k}\big(1\!-\! H_\ell(h|x)\,\tilde g_\ell\big)\big|\{x\ge h\}\Big] \Big)\right\rangle_{\{K_\ell\}}\ .
\label{pi}
\eea 
for the limiting joint distribution of single-site percolation probabilities $g_i$ and single site precisions $\omega_i$. In this equation, we once more have $x \sim \cN(0,1/\Omega_k)$.

\section{Results}
Here we collect a couple of results obtained for the Gaussian level-set percolation problem for random graphs in the configuration model class. Results can of course only indicate general trends and not be exhaustive, given the countless variations that could be contemplated.

\begin{figure}[b!]
  \begin{center}
   \includegraphics[width = 0.475\textwidth]{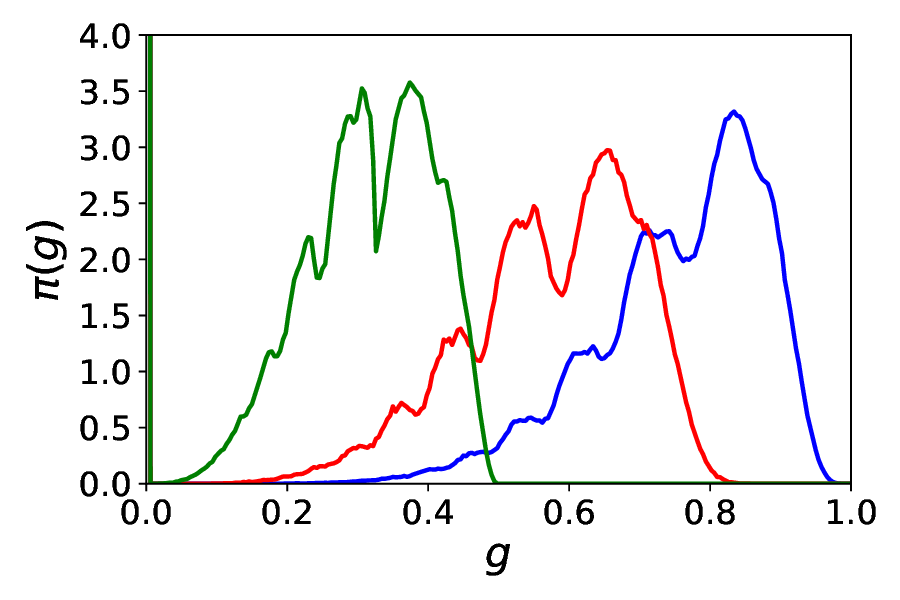}\hfil 
   \includegraphics[width = 0.475\textwidth]{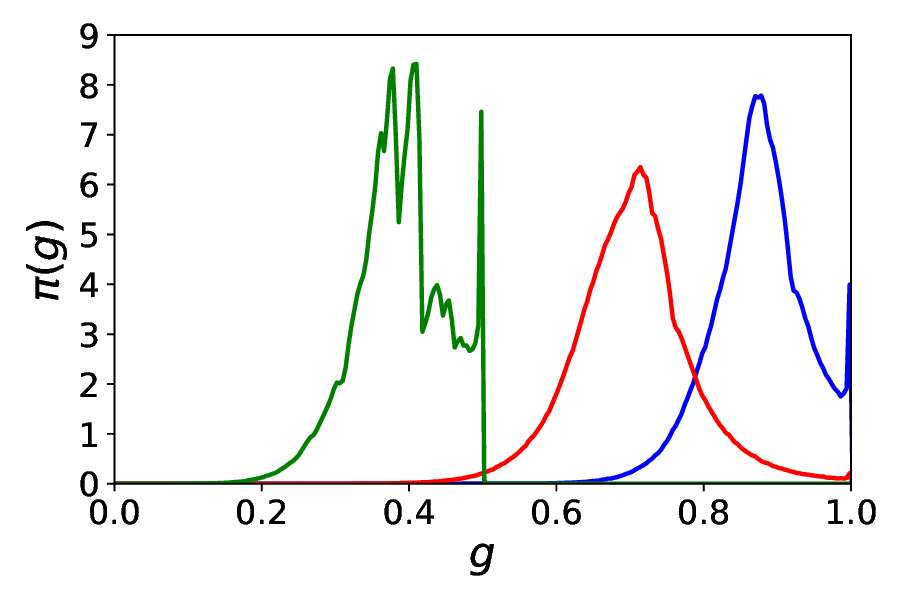}
  \end{center}
  \caption{(color online)  Left panel: Distribution $\pi(g)$ of local percolation probabilities at levels $h=-1$ (blue, rightmost curve), $h=-0.5$ (red, middle curve), and $h=0$ (green, leftmost curve) for an ER graph with mean degree 2, evaluated in the thermodynamic limit. (Right panel:) distributions for the same values of $h$ are displayed for a graph with power law degree distribution, $p_k \sim k^{-3}$ for $2\le k \le 125$.} 
\end{figure}

Figure 1 shows the distribution $\pi(g)$ for the massless Gaussian field at levels $h=-1$, $h=-0.5$ and $h = 0$ on an ER graph of mean degree $\langle k\rangle = 2$ and for a system with power law degree distribution $p_k \sim k^{-3}$ for $2\le k \le 125$ in the thermodynamic limit. In the first case, the original graph contains finite components, generating a $\delta$-peak at 0 in $\pi(g)$, whereas in the latter it doesn't. In both cases, the center of mass, i.e. the average percolation probability of the distributions decreases with increasing value of $h$ as expected. However, the shape of the distributions also changes markedly with the level $h$, thus carrying information that goes far beyond the respective average percolation probabilities. 

\begin{figure}[h!]
\setlength{\unitlength}{1mm}
  \begin{picture}(165,65)(5, 5)
   \put(5,5){\includegraphics[width = 0.475\textwidth]{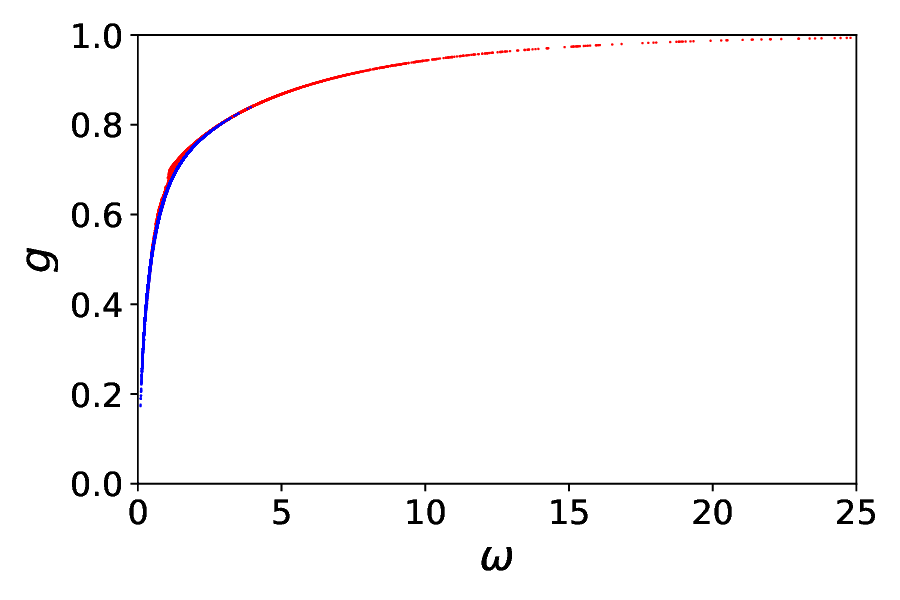}}
   \put(27.5,16.5){\includegraphics[width = 0.30\textwidth]{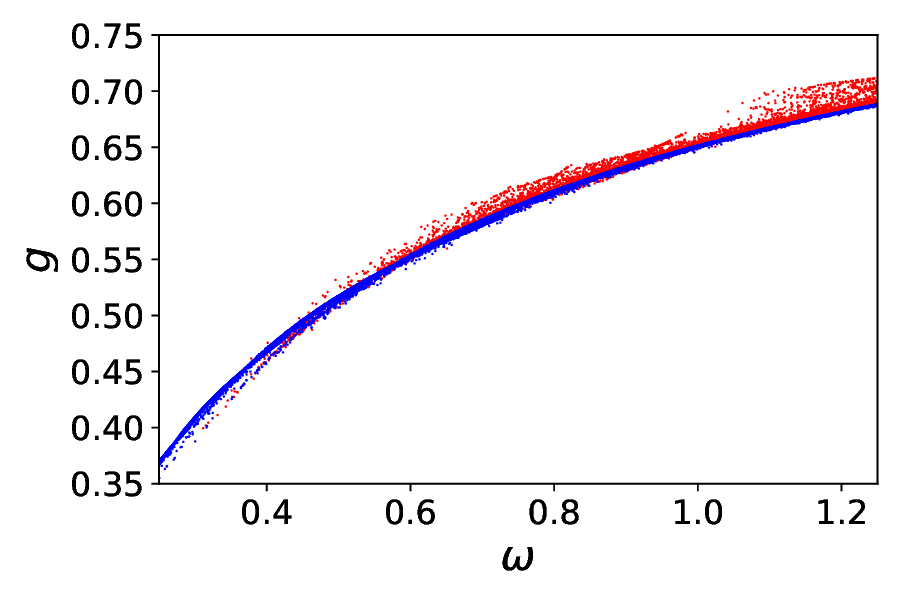}}
   \put(88,5){\includegraphics[width = 0.475\textwidth]{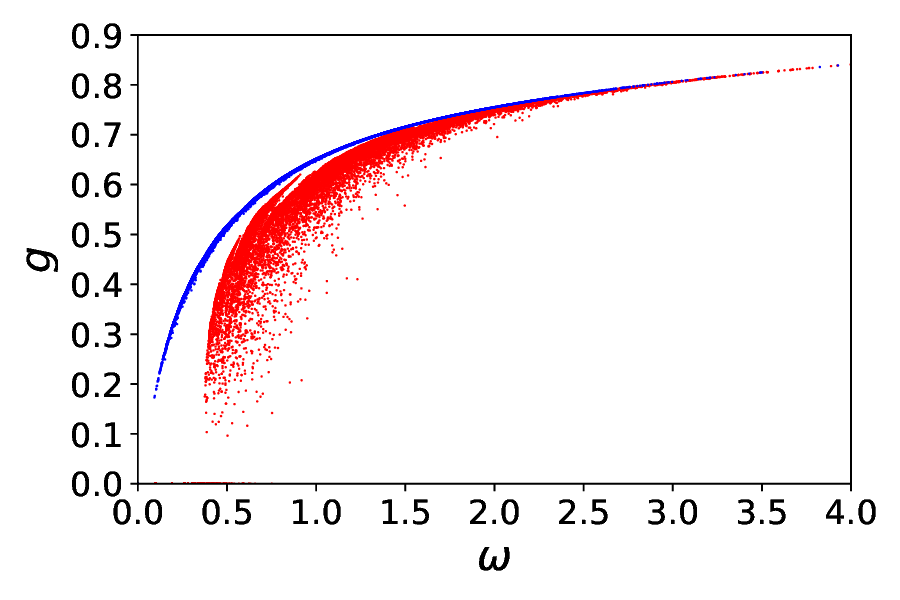}}
  \end{picture}
\caption{(color online) (Left panel:) Scatterplot of local level set percolation probabilities $g_i$ at levels $h=-0.5$ against local single node precisions $\omega_i$ for a massless Gaussian field defined on an ER graph with mean degree 2 of size $N=100,000$ (blue dots), and for a a graph of the same size with power law degree distribution $p_k \sim k^{-3}$ for $2\le k \le 125$ (red dots). The inset shows a zoom into the region of small $\omega$ region. (Right panel:) The same data for the massless Gaussian field on the ER graph of mean degree 2 (blue dots) are displayed together with data obtained for a system of the same type, but now for a Gaussian field with mass parameter $\mu=0.1$ (red dots).}
\end{figure}

In Fig.\,2 we display a scatterplot of marginal level-set percolation probabilities $g_i$ versus marginal Gaussian precisions $\omega_i$ for the same systems. The results suggest that --- remarkably --- the $g_i$ are up to very small uncertainties {\em determined\/} by the $\omega_i$, and the relation between the two appears to be even insensitive to the underlying graph structure, with the curve for the graph with power-law distributed degrees overlapping with that for the ER graph, but extending it to larger $\omega$ and $g$ values.  The inset shows that there is, however, some dispersion of the $g_i$ at given $\omega$ which is a bit more pronounced for the graph with power-law distributed degrees, so the $\omega_i$ cannot be taken as exact predictors. The figure also shows that the nearly perfect correlation between the $g_i$ and the $\omega_i$ disappears if the Gaussian field acquires a non-zero mass $\mu >0$.

\section{Random Regular Graphs}
It is easy to convince oneself that in the case of random regular graphs with degree distribution $p_k= \delta_{k,c}$ and uniform couplings Eq.\,\eqref{tpi} is self-consistently solved by $\tilde\pi(\tilde g,\tilde \omega) = \delta(\tilde g - \tilde g_0) \delta(\tilde \omega - \tilde \omega_0)$, where $\tilde\omega_0$ must be a solution of Eq.\,(13) of the letter, i.e., $\tilde\omega_0 =\tilde\omega_+$, and $\tilde g_0$ must be a solution of Eq.\,(15) of the letter, respectively.

This allows one to obtain percolation probabilities as a function of the level $h$. An example for 3 different mean degrees is shown in Fig.\,2 of the letter. In Fig.\,3 below we reproduce results of the stability analysis that produce the $c$ dependent critical levels $h_c(c)$ for  for a range of degrees ranging from 3 to 20.

\begin{figure}[h!]
    \begin{center}
    \includegraphics[width = 0.6\textwidth]{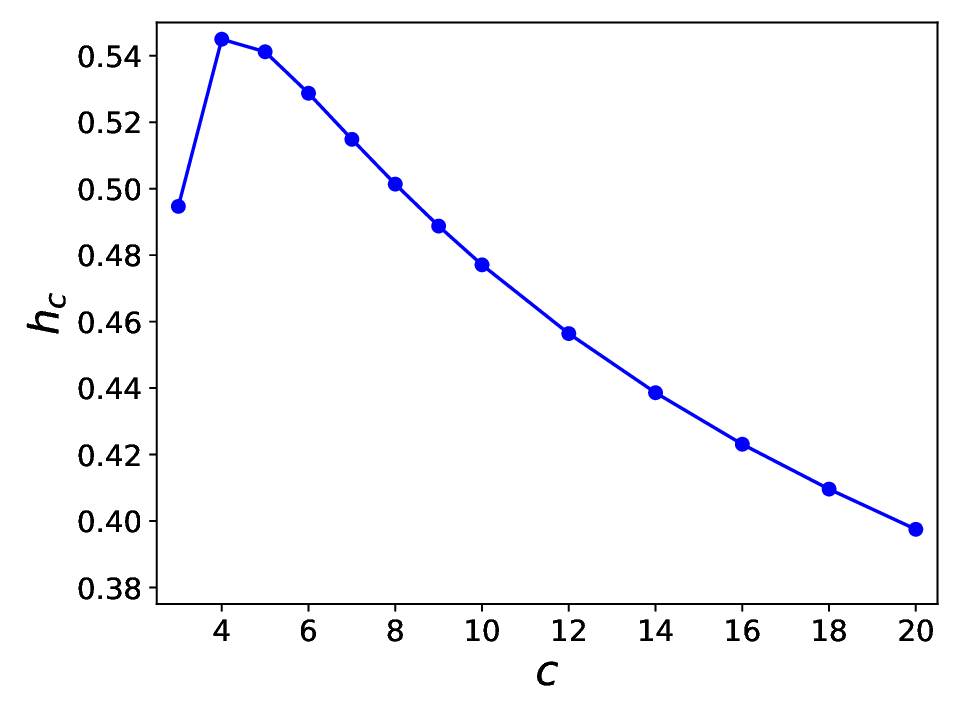}
    \end{center}
    \caption{Critical levels $h_c$ for Gaussian level-set percolation on $c$RRGs with $K=1$ and $\mu=0$ as a function of $c$, showing non-monotonic behaviour at small $c$. The line is a guide to the eye.}
  \end{figure}

\section{Cavity Equations in the Vicinity of the Percolation Transition}
For $h\lesssim h_c$ it is expected that the cavity probabilities $g_j^{(i)}$, hence the site-dependent percolation probabilities $g_i$ will be small. Following \cite{KuRog17}, we assume an expansion of the $g_j^{(i)}$ in Eqs.\,(10) of the letter in powers of $0\le \varepsilon = h_c-h \ll 1$ of the form
\be
g_j^{(i)} = \varepsilon a_j^{(i)} + \varepsilon^2 b_j^{(i)} + \varepsilon^3 c_j^{(i)} + \dots
\ee
and we expand the $H_\ell(h|x_j)$ appearing on the r.h.s. of these equations in a Taylor series at $h_c$,
\be
H_\ell(h|x_j) = H_\ell(h_c|x_j) + H'_\ell(h_c|x_j) (h-h_c) + \frac{1}{2}H''_\ell(h_c|x_j) (h-h_c)^2+ \dots \ .
\ee
Inserting these expansions into Eqs.\,(10) and collecting powers of $\varepsilon$ we obtain
\bea
\cO(\varepsilon): a_j^{(i)}\!\! &=&\!\! \sum_{\ell\in \partial_j\setminus i} B_{(ij),(j\ell)} a_\ell^{(j)}\label{O1}\\
\cO(\varepsilon^2): b_j^{(i)}\!\! &=&\!\! \sum_{\ell\in \partial_j\setminus i} B_{(ij),(j\ell)} b_\ell^{(j)} - \sum_{\ell\in \partial_j\setminus i} B'_{(ij),(j\ell)} a_\ell^{(j)}\nn\\
& & - \frac{1}{2}\bigg[\sum_{\ell,\ell'\in \partial_j\setminus i} B^{(2)}_{(ij),(j\ell)(j\ell')} a_\ell^{(j)} a_{\ell'}^{(j)} -\sum_{\ell\in \partial_j\setminus i} B^{(2)}_{(ij),(j\ell)(j\ell)} \big(a_\ell^{(j)}\big)^2\bigg]\label{O2}\\
\cO(\varepsilon^3): c_j^{(i)}\!\! &=&\!\! \sum_{\ell\in \partial_j\setminus i} B_{(ij),(j\ell)} c_\ell^{(j)} - \sum_{\ell\in \partial_j\setminus i} B'_{(ij),(j\ell)} b_\ell^{(j)} + \frac{1}{2}\sum_{\ell\in \partial_j\setminus i} B''_{(ij),(j\ell)} a_\ell^{(j)}\nn\\
& & - \bigg[\sum_{\ell,\ell'\in \partial_j\setminus i}\!\!\!\! B^{(2)}_{(ij),(j\ell)(j\ell')} a_\ell^{(j)} b_{\ell'}^{(j)}
-\sum_{\ell \in \partial_j\setminus i}\!\!\! B^{(2)}_{(ij),(j\ell)(j\ell)} a_\ell^{(j)} b_{\ell}^{(j)}\bigg] 
+ \frac{1}{3!}\bigg[\sum_{\ell,\ell',\ell''\in \partial_j\setminus i}\!\!\!\! B^{(3)}_{(ij),(j\ell)(j\ell')} a_\ell^{(j)} a_{\ell'}^{(j)} a_{\ell''}^{(j)}\nn\\
& & -3 \sum_{\ell,\ell'\in \partial_j\setminus i} B^{(3)}_{(ij),(j\ell)(j\ell')(j\ell')} a_\ell^{(j)}\big(a_{\ell'}^{(j)}\big)^2 + 2 \sum_{\ell\in \partial_j\setminus i} B^{(3)}_{(ij),(j\ell)(j\ell)(j\ell)} \big(a_{\ell}^{(j)}\big)^3\bigg]
\label{O3}
\eea
in which 
\be
B^{(2)}_{(ij),(j\ell)(j\ell')} = \E_{x_j}\big[H_\ell(h|x_j) H_{\ell'}(h|x_j)\Big|\{x_j\ge h\}\big]\Big|_{h_c}\ .
\ee
and
\be
B^{(3)}_{(ij),(j\ell)(j\ell')(j\ell'')} = \E_{x_j}\big[H_\ell(h|x_j) H_{\ell'}(h|x_j)H_{\ell''}(h|x_j)\Big|\{x_j\ge h\}\big]\Big|_{h_c}\ .
\ee
are non-zero elements of second and third order generalizations of the non-backtracking operator defined in Eq.\,(11) of the letter, while the $B'_{(ij),(j\ell)}$ and the $B''_{(ij),(j\ell)}$ are first and second $h$-derivatives of the non-zero elements of $B$, all evaluated at $h_c$. 

Following \cite{KuRog17}, we can obtain a more compact representation of the above relations by introducing a $2M=\sum_i k_i$-dimensional vector $\va =\big(a_j^{(i)}\big)$. Equation \eqref{O1} then states that $\va$ is the Frobenius (right)-eigenvector of $B$ corresponding to the eigenvalue $\lambda_{\rm max}(B) = 1$. Setting  $\va = \alpha \vv$ with $\vv$ normalized to $||\vv||_1=1$ and an amplitude $\alpha$ to be determined and denoting by $\vu^T$ the corresponding Frobenius (left) eigenvector of $B$, i.e. $\vu^T = \vu^T B$, with $\vu^T \vv =1$, Eq.\,\eqref{O2} yields
\be
\alpha = - 2\frac{\vu^T B'\vv}{\vu^T B^{(2)} \vv\otimes\vv - \vu^T \tilde B^{(2)} (\vv \vv)}
\label{alpha}
\ee
Here we use the notation $(\va \vb)$ to denote a vector with components given by the product of components of $\va$ and $\vb$, i.e., $(\va \vb)_\ell^{(j)} = a_\ell^{(j)} b_\ell^{(j)}$ and an analogous construction for a vector $(\va \vb \vc)$ constructed from three vectors, and $\tilde B^{(2)}$ to denote the matrix obtained by restricting $B^{(2)}$ to be diagonal in the second pair of indices.

In order to obtain the $\cO(\varepsilon^2)$ contribution to the $g_j^{(i)}$ we rewrite Eq.\,\eqref{O2} as
\be
(\bbbone - B)\vb = - \alpha B'\vv - \frac{\alpha^2}{2}\Big(B^{(2)}\vv\otimes\vv -\tilde B^{(2)}(\vv\vv)\Big)\ .
\ee
It is solved by
\be 
\vb = (\bbbone - B)^+ \bigg[- \alpha B'\vv - \frac{\alpha^2}{2}\Big(B^{(2)}\vv\otimes\vv -\tilde B^{(2)}(\vv\vv)\Big)\bigg] + \beta \vv\ ,
\label{defw}
\ee
in which $(\bbbone - B)^+$ is the Moore-Penrose pseudoinverse of $(\bbbone - B)$ and $\beta$ a coefficient to be determined from Eq.\,\eqref{O3}. Indeed, writing Eq.\,\eqref{defw} as $\vb = \vw + \beta \vv$, thereby defining $\vw$, multiplying Eq.\,\eqref{O3} by $\vu^T$ yields
\bea
0 &=& -\vu^T B'\vw +\frac{\alpha}{2}\vu^T B''\vv 
    -\alpha\Big[\vu^T B^{(2)}\vv\otimes \vw - \vu^T\tilde B^{(2)}(\vv\vw)\Big]\nn\\
& & +\frac{\alpha^3}{3!}\Big[\vu^T B^{(3)}\vv\otimes\vv\otimes\vv - 3\vu^T \tilde B^{(3)}\vv\otimes(\vv\vv)
    + 2\vu^T \tilde{\tilde B}^{(3)}(\vv\vv\vv)\Big]\nn\\
& & +\beta\Big[-\vu^T B'\vv -\alpha \vu^T B^{(2)}\vv\otimes \vv + \alpha \vu^T \tilde B^{(2)}(\vv\vv)\Big]\ ,
\eea
which is easily solved for $\beta$, thereby completely determining $\vb$. (Note that Eq.\,\eqref{alpha}) can be used to simplify the coefficient of $\beta$ in the last equation to $\vu^T B'\vv$).
\bibliography{GLSSupp.bbl}